\documentclass[aps,twocolumn,preprintnumbers,amsmath,amssymb,floatfix,groupedaddress]{revtex4}

\usepackage{enumerate}
\usepackage{graphicx}
\usepackage{color}
\usepackage[utf8]{inputenc}
\usepackage{enumitem}

\usepackage{fontenc}  
\usepackage{amsmath}

\begin{document}

\title{Revisiting Quantum Mysteries.} 

\author{Philippe Grangier}

\affiliation{ %\vskip 2mm
Laboratoire Charles Fabry, Institut d'Optique Graduate School,  CNRS,
Universit\'e Paris~Saclay, F~91127 Palaiseau, France, \email{philippe.grangier@institutoptique.fr}}

\begin{abstract}
In this article we argue that in quantum mechanics, and in opposition to classical physics,  it is impossible to say that an isolated  quantum system ``owns" a physical property. Some properties of the system, its mass for example, belong to it in a sense close to that of classical physics; but most often a property must be attributed to the system  {\it within a context.} We give simple motivations for adopting this point of view, and show that it clarifies many issues in quantum physics. 

\end{abstract}

\maketitle

\section{Our quantum world. }

In many venues intended for the general public one encounters the assertions that in the ``quantum world" an object can be in several places at the same time, or that two particles located at astronomical distances can influence each other instantaneously, or that particles pass through barriers, and so forth. 

However, the world thus described is nothing but the one we live in. Those who are familiar with the popularization of quantum physics are used to hearing these sentences, which fascinate and challenge. They often lead to the ideas, quite frustrating for the uninitiated, that quantum mechanics is a discipline full of paradoxes, and incomprehensible otherwise than by its mathematical formalism; and  that even the best physicists disagree on what it means  \cite{Laloe,Peres}.

Given that quantum mechanics provides the basis for innumerable technologies (such as electronics, lasers, or medical imaging) which have changed our daily life, and has never been disproved as a theory, this difficulty in communicating its physical content to the layman is surprising. It may even appear embarrassing, in a context where many countries are investing billions of Euros to develop technologies related to the ``second quantum revolution",  associated with new ways to process information, but even more difficult to explain than the first one.
%\\

A legitimate question is then : Would it be possible to formulate different statements, referring specifically to this mysterious ``quantum world", without appearing as a series of contradictions or absurdities? This is what we propose here, also avoiding  any excessive recourse to mathematics: it is indispensable, but will only be  briefly mentioned in the last part of the article. 

\vspace{-3mm}
\section{Empirical evidence and its consequences.}
\vspace{-1mm}

Our starting point will be a series of empirical observations, that is to say, observations that relate to experiments that are quite feasible and that have been carried out, some of them recently. We will start from concepts known at the beginning of the 20th century (such as light is a wave, or matter is made of atoms with a nucleus and electrons) and we will introduce new phenomena, which we will then explain. 
\\

The fundamental physical fact is that observations of microscopic objects show a discrete, or quantized, character. Here are some examples: 
\begin{itemize}

\item  The possible wavelengths of light emitted by atoms take only certain particular values, with each atom characterized by the ``spectrum" of wavelengths that it can emit.

\item   Light can extract electrons (particles) from matter, but in order  to describe this phenomenon, called the photoelectric effect, we are led to admit that light is also formed of particles (photons).

\item   We can perform experiments where objects classically considered as particles (electrons, atoms...) give rise to interference effects, which are classically associated with waves, and not with particles.

\end{itemize}
A rather inevitable conclusion of these observations is that microscopic objects (atoms, electrons, photons...) have behaviors that combine a discrete character (i.e. described by integers) with the possibility of interference (traditionally associated with continuous waves).

Quantum mechanics was built, on the basis of these observations, by a few brilliant physicists who invented a somewhat bizarre mathematical formalism, able to account for all these observations, and also to make countless predictions of new phenomena. This formalism, or even these formalisms, because there are several equivalent ones, have never been put in default, but the physical nature of the objects and properties they describe has remained obscure. The questioning on how to describe (or even define) the ``physical reality", quite legitimate in our view,  has given rise to innumerable interpretations and paradoxical statements - the sentences quoted at the beginning of this article are a very  limited sample. 

It is therefore clear that an ``ingredient" is missing that would give a physical meaning to this very efficient formalism;  this is what we propose in the following. 

\vspace{-3mm}
\section{Underestimating empirical evidence ? }
 
The normal approach in physics since Newton has been  to define objects, to attribute properties to them, and to measure these properties. One then asserts that the object ``has" this property, for example that it has a position, a velocity, or a momentum. Does this ``natural" approach work in quantum mechanics? Although physicists are extremely reluctant to admit it, the answer is clearly no - and this ``no", correctly interpreted, provides the empirical element that is missing in our understanding of the quantum description of the physical world. 

So, let's look at a simple experiment that shows that it is impossible to say that a quantum particle ``owns" a property. Of course some particle's properties, its mass for example, belong to it in a sense very close to classical physics; but in general the situation is not so simple.
One can for example measure the angular momentum of an atom along a certain direction $\bf u$, or the polarization (equivalent to angular momentum) for a photon. Such a measurement gives a discrete result, as explained above, and we can therefore say for example that we obtain the value +1, in adequate units. Moreover, if we repeat immediately (ideally instantly) this same measurement on the same particle, we find +1 again: the result can therefore be predicted with certainty, as one would expect in classical physics. 

But let's complicate things a bit, and measure the angular momentum in another direction, let's say $\bf v$. We find for example +2, then we come back to $\bf u$, and we measure again, on the same particle. In a classical framework we would expect to find the previous +1 result, but in fact we don't necessarily get +1, but a random result! In general this new result can take any value in a certain set defined by the system studied, for example (-2, -1, 0, +1, +2) for a certain type of atom, each of these values having a certain probability of being obtained. One sometimes speaks of ``measurement-induced perturbation", a phenomenon that also exists in classical physics; but the randomness described here is inescapable: it is not possible to eliminate it completely, nor to trace it back to a ``true value" of the measured quantity. 

Very often this simple but fundamental experiment is not considered as a fact, but is immediately buried in mathematical formalism, invoking operators that do not commute or other arguments that are not explanations, but mathematical descriptions. However, this experiment is essential since it shows that the value attributed to certain physical properties varies in an irremediably random way when the measurement process is modified. However, this is not an incomprehensible blur, since as long as the same measurement is repeated on the same particle, the measured value always remains the same: certainty and repeatability remain, but under more restrictive conditions than those observed in classical physics. 

If one legitimately considers that certainty and reproducibility are minimum requirements for a realistic description of the physical world, the inescapable consequence of the above observations is that the object to which one must attribute physical properties is not ``a system", but ``a system on which a given measurement is made", since only in this case the result does not change. Such a ``contextual" description is an essential difference from Newtonian physics, and it can be used as a basis to build up  quantum physics, as we will now see. 

\section{Contexts, systems and modalities. }

These observations being very quantum but not really mysterious, we can formalize them (still without mathematics!) by introducing some definitions \cite{CO2002,csm1,csm4b,completing,inference,debate} : 
\begin{itemize}

\item   let us designate the devices, arranged to carry out a particular measurement on a given quantum system, by the word ``context";

\item   let us refer to a particular measurement result, on a given system in a given context, by the word ``modality".  By definition a  modality characterizes the system as well as  it is physically possible, and, once obtained, it is ideally repeatable with certainty as long as the context is not changed; 

\item   let us admit the following fundamental principle, called ``contextual quantization": {\it whatever the context, a measurement on a given system gives one modality among $N$ possible ones, where the value of $N$ characterizes the system; these $N$ modalities are mutually exclusive, i.e. only one can be realized at a time.}

\end{itemize}

It  follows from the above that if there are other different modalities in other contexts, the link between the modalities of different contexts can only be probabilistic; otherwise one would have a ``super context" with a number of mutually exclusive modalities greater than $N$, which would contradict the fundamental principle \cite{csm4b}. 

In this approach the probabilities in quantum mechanics are thus physically necessary, since they are imposed by the contextual quantization. It is crucial to note that {\bf the conjunction of the two ingredients, quantization and contextuality, is required}. For example, many probabilistic phenomena are discrete but classical (heads or tails, dice game, and so forth), and others have contextual aspects (for example in a poll it is known that the order of the questions influences the answers) but they remain classical because quantization does not play a constraining role. 

\section{A little bit of mathematics. }
\vspace{-2mm}

The last step of our approach is to mathematize the principle of contextual quantization, to obtain the quantum formalism. It is crucial to emphasize that, as usual in physics, this step cannot be ``deductive" (we cannot show that the formalism is the only one possible), but must be ``inductive"~: we propose a formalism, the most general one possible, respecting the imposed physical conditions, and we verify deductively that it works by describing well the observed effects.  The theoretical question is therefore~: given a modality (among $N$) in an initial context, resulting from a previous measurement, what is the probability of obtaining another modality (among $N$) in a final context associated with a new measurement ? We are thus looking for an $N \times N$ matrix of probabilities between the initial context and the final context, which is generally called a stochastic matrix. 

The fundamental (and actually unique) mathematical ingredient is then to associate a projector ($N \times N$ matrix) with the set of all modalities that are related to each other with certainty, either in the same context (it is then the same modality) or in different contexts (this link by ``transfer of certainty"  is observed empirically). This set constitutes an equivalence class of modalities that are linked ``extracontextually", and it is therefore called an extravalence class. 

By admitting this fundamentally quantum idea, and a few simple consistency arguments \cite{csm4b}, there is in fact no longer any choice: powerful mathematical theorems allow us to show that the only possible theory is quantum mechanics. 
More precisely, Uhlhorn's theorem \cite{uhlhorn,chevalier} shows that unitary transformations between projectors are necessary to preserve the mutually exclusive character of events in each context; and Gleason's theorem \cite{gleason,cooke} shows that Born's  rule is necessary to respect the general structure of a probability law (see details in the Appendix).  

The probability matrix we are looking for has then a particular form: it is unistochastic, i.e. it is constituted by the squared moduli of the coefficients of a unitary matrix. It also follows that a context is associated to a family of $N$ orthogonal projectors, and that a change of context is associated to a unitary transformation between these families, which reconstitutes the usual framework of quantum probabilities \cite{csm4b}. Within this framework,  a usual quantum state $| \psi  \rangle$ is predictively incomplete, because it is associated to an extravalence class, and not to a single modality;  the specification of the context is needed to define a true probability distribution \cite{completing,inference}, which is consistent with attributing a modality to a system within a context.

\section{No paradox anymore ?}
\vspace{-2mm}

 We will not go further into the mathematical details here, but to conclude let us try to translate into this new language the paradoxical statements quoted at the beginning of this article, again avoiding any formalism: 

\begin{itemize}

\item {\it an object can be in several places at the same time:} we often hear also that a quantum bit is in several states at once, and that this corresponds to the notion of quantum superposition. But what does this ``at the same time" mean? In fact, it is a question of detecting a modality, therefore a certainty, but a certainty in another context. To say that a particle is ``at the same time" in the two arms of the interferometer means in fact that if these two arms interfere constructively, we can predict with certainty where the particle will go. On the other hand, if the measurement is made inside the interferometer, then the result is random, according to the principles explained above: the measurements either inside or outside the interferometer correspond to different contexts. 
\vskip 2mm

\item  {\it two particles separated by an arbitrarily large distance can influence each other instantaneously:} this is the phenomenon of entanglement, and there is in fact no instantaneous influence at a distance, but again the manifestation of a modality, which is a certain and reproducible property of the pair of particles, e.g. that the total angular momentum of the pair is zero. This can be verified in the appropriate context (by making a joint measurement on both particles), but if one performs the measurement in a context where the particles are spatially separated one will obtain necessarily random results (because of the change of context, see above).  Still  Born's rule must holds when the context is changed, and for appropriate measurements on entangled particles, it predicts  strongly correlated results. This is often related to ``quantum non-locality", but in our framework it rather  illustrates the predictive incompleteness of $ |\psi \rangle$ without a context \cite{inference}.
\vskip 2mm

\item {\it the quantum particles pass through the barriers:} this is the tunneling effect, an ``obvious" effect for classical waves (it is due to so-called evanescent waves), but ``impossible" for classical particles. But the classical notions of waves and particles are inadequate here, and one must consider probability amplitudes, which allow one to get a physical picture of Born's law in this case. The change of context corresponds then to the passage from a representation where the momentum of the particle is defined, to another ``incompatible" one where its position is defined. A usual quantum wave-packet is somewhere between these two extremes,  but in any case speaking about particles or waves with a classical behavior is a fiction, sometimes useful but most often misleading. 
\vskip 2mm

\end{itemize}

The above examples do not necessarily require to fully determine the modalities, i.e. all the system's properties, but the general ideas remain valid. It may be noticed that contextual quantization is difficult to admit, even for physicists, since it asserts that properties/modalities belong to a system within a context, and not to a system considered as ``alone in the universe", as classical physics would assume. But the notion in itself is not paradoxical, and if we admit that such is the behavior of the nature in which we live, then the statements related to the ``quantum world" finally cease to appear as a long series of contradictions or absurdities.

\vspace{-3mm}
\section{Quantum computing in context.}

This point of view also underlines that statements made ``out of context" (such as those concerning the famous Schr\"odinger's cat, as being both dead and alive) are meaningless. Any statement concerning a quantum system, even a large one such as a quantum computer, is only meaningful in combination with a relevant context in which the modalities under consideration can manifest themselves and be observed. 

This allows us to come back to the technological promises mentioned at the beginning of this article. It is true that a quantum computer has, in principle, a gigantic computing power, based on the manipulation of entangled states involving a very large number of quantum bits. But this manipulation will only be possible and meaningful if an adequate context is available, which will also be of a very high complexity - even if it is then a classical complexity. 

Then is it certain that the ensemble formed by the quantum ``heart" (the system) surrounded by classical instruments capable of interacting with it (the context) will be more efficient than a usual, classical supercomputer ? The answer to this question remains open,  and the scientific and rational conclusion is that we have to go forward and see what happens - it may not be exactly what we expect, but this is very common in experimental physics, and it could be just as interesting. 

\section*{Appendix : Uhlhorn's and Gleason's theorems.}

As stated above, the empirical facts that we want to describe mathematically are :
\vskip 2mm

\noindent {\bf (i)}  in each context a measurement provides one modality among $N$ possible ones, that are mutually exclusive. No measurement can provide more than $N$ mutually exclusive modalities, and once obtained in a given context, a modality corresponds to a certain and repeatable result, as long as one remains in this same context. 
\vskip 2mm

\noindent { \bf (ii)}  the certainty and repeatability of a modality can be transferred between contexts; this fundamental property  is called  extracontextuality of modalities. All the modalities related together  with certainty, either in the same or in different contexts, constitute an equivalence class that we call an extravalence class. 
\vskip 2mm

Our central mathematical ingredient is then to associate a rank-one projector $P_i$ (a $N \times N$ hermitian matrix such as $P^2 = P$ ) to each extravalence class of modalities, in such a way that the $N$ projectors associated with  the $N$ modalities within a given context are mutually orthogonal. In addition, we assume  that the probability to get a given modality is a function of its associated projector only.  The heuristic motivation for this choice is that it ensures that the events associated with modalities cannot be subdivided in more elementary events, as it would be the case with classical (partition-based) probabilities. 

The choice of a specific orthogonal set of projectors associated with a context is not given a priori, but mutually exclusive modalities in the context should stay so, whatever choice is made for the projectors. It means that if two orthogonal projectors are associated with two mutually exclusive modalities, they should stay orthogonal whatever choice is made for the projectors associated with a ``reference" (fiduciary) context.  Then Uhlhorn's theorem \cite{uhlhorn,chevalier} warrants that the transformations between the sets of projectors associated  with different contexts must be unitary or anti-unitary;  anti-unitary operators are associated with time reversal, so we will omit them here and consider only unitary ones.  

We thus get a major result : once a set of mutually orthogonal projectors associated with a fiduciary context has been chosen, the sets of projectors associated to all other contexts are obtained by unitary transformations, so we are simply ``moving" in a Hilbert space.  
There are also various arguments for using unitary (complex) rather than orthogonal (real) matrices; in our framework the simplest argument is to require that all permutations of modalities within a context are continuously connected to the identity, which is not possible with orthogonal  matrices, but is possible with unitary ones.

The next step is to consider the probability $f(P_i)$ to get a modality associated with projector $P_i$.  By construction a context is such that $\sum_{i=1}^{i=N}  P_i = I$, and $\sum_{i=1}^{i=N}  f(P_i) = 1$ for any complete set $\{ P_i \}$.  But these are just the assumptions of Gleason's theorem \cite{gleason,cooke}, so there  is a density matrix $\rho$ such that $f(P_i) = \textrm{Trace}(\rho P_i)$. If the value 1 is reached, then  $\rho$ is also a projector $Q_j$ and $f(P_i) = \textrm{Trace}(Q_j  P_i )$ which is the usual Born's formula. 

We have thus reconstructed the basic probabilistic framework  of quantum mechanics. In order to use it, one must define explicitly the relevant physical properties and associated contexts, that may go from space-time symmetries (Galileo group, Lorentz group) \cite{book} to qubits registers. In any case, contextual quantization applies and sets the scene where the actual physics takes place. 

An interesting question is whether this probabilistic formalism might be used outside physics, for its own virtues in a ``quantum-like" analogy \cite{AK1}. %,AK2}. 
However, it can be observed that the requirements (i) and (ii) written above are strictly respected in quantum mechanics, hence the strong predictive power, whereas this is only loosely the case in other domains.  Therefore whether or not a variant of this formalism might  have a useful predictive (and not only descriptive) power outside physics remains an open - and quite stimulating - question. 

{\bf Acknowledgements. } 
The author thanks Alexia Auff\`eves, Roger Balian, Nayla Farouki and Franck Lalo\"e for many interesting and stimulating discussions, 
and Arkady Plotnitsky for quite useful remarks.

\end{document}